\documentclass[pra,twocolumn,amsmath,amssymb]{revtex4}
\usepackage{amsmath}
\usepackage[pdftex]{graphicx}

\usepackage{bm,ulem}

\newcommand{\beq}{\begin{eqnarray}}
\newcommand{\eeq}{\end{eqnarray}}
\newcommand{\be}{\begin{equation}}
\newcommand{\ee}{\end{equation}}

\def\H1{\widehat{H}_1}

\newcommand*{\bfrac}[2]{\genfrac{}{}{0pt}{1}{#1}{#2}}
\def \av#1{{\langle#1\rangle}}
\def \ket#1{{|#1\rangle}}

\def \braket#1#2{{\langle #1|#2 \rangle }}

\begin{document}

\title{Non-equilibrium dynamics of the Tavis-Cummings model}
\author{Peter Barmettler$^{1}$, Davide Fioretto$^{2,3}$, Vladimir Gritsev$^{2}$}
\affiliation{$^{1}$ Department of Theoretical Physics,
 University of Geneva, 24, Quai Ernest-Ansermet, 1211 Gen\`{e}ve 4, Switzerland\\
$^{2}$ Physics Department, University of Fribourg, Chemin du
Mus\'ee 3, 1700 Fribourg, Switzerland\\
$^{3}$ Institute for Theoretical Physics, University of Amsterdam,
Science Park 904, Postbus 94485, 1090 GL Amsterdam, The Netherlands}

\begin{abstract} 
In quantum many-body theory
no generic microscopic principle at the origin of complex dynamics is known. Quite opposed, in classical mechanics the theory of non-linear dynamics provides a detailed framework for the distinction between near-integrable and chaotic systems. Here we propose to describe the off-equilibrium dynamics of the Tavis-Cummings model by an underlying classical Hamiltonian system, which can be analyzed using the powerful tools of
classical theory of motion. We show that 
scattering in the classical phase space can drive the quantum model close to thermal equilibrium.
Interestingly, this happens in the fully quantum regime, where physical
observables do not show any dynamic chaotic behavior.
\end{abstract}

\maketitle
Many aspects of the transition from
regular dynamics of an integrable system to erratic behavior of a
complex system are understood in classical mechanics. On the one hand,
there is the Kolmogorov-Arnold-Moser (KAM) theorem \cite{KAM}, that proves the stability of
weakly perturbed integrable systems. On the other hand, a variety of mechanisms
leading to chaos and eventually to the ergodic
exploration of phase space have been found (See e.g. \cite{Chaos}). For quantum systems, there exist quasiclassical \cite{Gutzwiller} and phenomenological (e.g. random matrices \cite{haake,Gutzwiller}) descriptions of quantum chaotic phenomena, but there remains an important conceptual gap
between regular and complex behaviors. In this Letter we investigate a non-trivial integrable quantum system away from 
the quasiclassical limit and 
gain microscopic insight into the emergence of irregularity when breaking integrability by driving an internal parameter.

A good starting point to approach regular dynamics of non-trivial quantum systems
are Bethe ansatz (BA) integrable models, which possess a complete set of integrals of motion. 
The exact solutions of time-independent BA many-body solvable systems
played a crucial role in the understanding of various fundamental phenomena and concepts in
physics. Famous examples are the solutions for the Ising model, the Heisenberg spin
chain, the one-dimensional Hubbard model or the Lieb-Liniger gas \cite{KorepinBook,Gaudin}.
Also certain aspects of quantum chromodynamics can be described by the integrable quantum
spin chain with complex spin \cite{QCD}.
However, the non-equilibrium dynamics of these models are rich \cite{Manmana2007,Barmettler2009,Faribault2009,FS13,MosselCaux2012,Iyer2012,Pozsgay2013,Fagotti2013,Kormos2013,CE13} 
and much more difficult to be calculated within the BA than the static properties.
Formulating a theory of integrability breaking for time-dependent problems is thus not only a conceptual, but also
a technical challenge.

The main finding presented in this Letter is that there exists an exact description of the quantum Tavis-Cummings model in terms of a classical many-body interacting system. Deviation from integrability
for the quantum system can then be understood in terms of the classical system, 
for which powerful tools such as the KAM theorem are available. 
It is important to note that this representation is not connected to the quasiclassical 
limit of the related Dicke model \cite{Emary2003}, but exact for the full quantum model. 
We will derive this classical representation and analyze its dynamics under periodic driving of the detuning.

The Tavis-Cummings model was introduced in the context of interaction of light and matter in quantum optics \cite{Tavis-Cummings}. It can be seen as a Dicke model \cite{Dicke-original} in the rotating wave approximation.
An application is for instance the description of the Bose-Einstein condensate 
in an optical cavity \cite{dicke1}. For our theoretical purposes, we take it as a representative of the class of homogeneous Gaudin models \cite{Ortiz2005}, which can be seen as minimal non-trivially BA integrable systems. Our results can be straightforwardly applied to arbitrary homogeneous Gaudin models, such as the Lipkin-Meshkov-Glick model for phase transitions in nuclei \cite{LMG}. The challenges one faces when extending our approach to inhomogeneous models, for instance the Richardson or the central spin model, will be discussed in the conclusions.

The quantum Hamiltonian of the Tavis-Cummings model reads
\begin{eqnarray}\label{eq:Dicke}
\hat H_{D}=\Delta \hat S^{z}+g(\hat b^{\dag}\hat S^{-}+\hat b \hat S^{+}),
\end{eqnarray}
where $\hat S^{\mu}=\sum_{j=1}^{2S}\frac{\hat \sigma_{j}^{\mu}}{2}$ is a collective spin
operator with $\sum_{\mu=1}^3 (\hat S^\mu)^2=S(S+1)$, the single-mode bosonic field
($\hat b$ and $\hat b^{\dag}$, photon annihilation and creation operators) is detuned by $\Delta$. The total number of
excitations $M=\hat b^{\dag}\hat b+\hat S^{z}+S$ is a conserved quantity. Therefore, the
relative strength of the detuning $\Delta/g$ is the
only free parameter in the system in a given sector with well defined quantum
numbers $M$ and $S$.

The Tavis-Cummings model belongs to the class Gaudin-type models \cite{Jurco} and one can introduce the Bethe wavefunction
\begin{eqnarray}
	|\{\lambda_\alpha\}\rangle=\prod_{\alpha=1}^{M}\hat B(\lambda_{\alpha})|0\rangle\, \label{eq:staticwave}\,.
\end{eqnarray}
The rapidities (or spectral parameters) $\lambda_\alpha$, $\alpha=1,\ldots, M$, are complex numbers, while the excitation creation operators $\hat B(\lambda)$ and the vacuum $|0\rangle$ are defined as
\begin{eqnarray}
\hat B(\lambda)=\hat b^{\dag}-g\frac{\hat S^{\dag}}{\lambda},\mbox{~and~}\,\hat b|0\rangle =\hat S^-|0\rangle=0.
\end{eqnarray}

The action of the Hamiltonian on the Bethe wave function is given by
\begin{eqnarray}
	\label{eq:hpsi}
\!\!\!\!\!\hat H |\{\lambda_\alpha \} \rangle\!&=&\!\!\left[E_{S,M}(\{\lambda_\alpha\})+\sum_{\alpha=1}^M f_\alpha(\{\lambda_\alpha\}) \right]\!\!\prod_{\alpha=1}^M B(\lambda_\alpha) |0\rangle\!\nonumber\\
&&\!\!\!\!\!\!\!\!\!\!+g\sum_{\alpha=1}^M \frac{f_\alpha (\{\lambda_\alpha \})}{\lambda_\alpha} \prod_{\beta \ne \alpha} B(\lambda_\beta) b^\dagger |0\rangle.
\end{eqnarray}

where $f_\alpha(\{\lambda_\alpha\})$ is defined as
\begin{eqnarray}
	f_\alpha(\{\lambda_\alpha\})=-\frac{2g^2S}{\lambda_{\alpha}}+\lambda_{\alpha}-\Delta+\sum_{\bfrac{\beta=1}{\beta\neq \alpha}}^{M}\frac{2 g^2}{\lambda_{\alpha}-\lambda_{\beta}}\,.
\end{eqnarray}
Eq. \ref{eq:hpsi} is known as the off-shell Bethe equation~\cite{Babujian}. If the rapidities satisfy the Bethe equations  $f_\alpha(\{\lambda_\alpha\})=0$, then the Bethe wavefunction is an eigenstate with the eigenenergy

\begin{align}
E_{S,M}(\{\lambda_\alpha\})=\Delta(M-S)-\sum_{\alpha=1}^{M}\lambda_{\alpha}\,.\label{eq:staticenergy}
\end{align}
Indeed, it is possible to construct a basis of Bethe states, and this is how the system is solved when the Hamiltonian is time-independent.
However, for the Tavis-Cummings model, one can explicitly include the off-shell term in order to describe the dynamics of the wavefunction under a time-dependent detuning $\Delta(t)$. The Bethe wave function including the off-shell part completely describes the solution of the time-dependent Schr\"odinger equation with rapidities $\{ \lambda_\alpha\}$) moving in time, 
\begin{eqnarray}
	|\Psi(t)\rangle=\exp[-ie(t)]\prod_{\alpha=1}^{M}\hat B(\lambda_{\alpha}(t))|0\rangle\,,\label{eq:dynamicwave}
\end{eqnarray}
with a phase $e(t)=\sum_{\alpha}\int_{0}^{t}[E_{S,M}(\lambda_{\alpha} (t))+f_{\alpha}(\lambda_{\alpha})]-S\Delta(t)$
and where the rapidities are subject
to the following set of equations
\begin{eqnarray}\label{lambda-eq}
&&i\frac{\dot{\lambda}_{\alpha}(t)}{\lambda_{\alpha}(t)}=f_\alpha\left(\lambda_\alpha (t)\right)\,. 
\end{eqnarray}
It can be verified that in the stationary case $\dot{\lambda}_{\alpha}(t)=0$ the time-dependent wave function \eqref{eq:dynamicwave} reduces to the static one \eqref{eq:staticwave} with a phase given by the eigenenergy \eqref{eq:staticenergy}. 

The appeal of the representation \eqref{lambda-eq} is its equivalence to a integrable classical many-body problem. After the change of variables
\begin{eqnarray}
\lambda_{\alpha}(t)=2 \,x^{2}_{\alpha},
\end{eqnarray}
the dynamical Bethe
equation \eqref{lambda-eq} reads
\begin{eqnarray}\label{eq:dicke-1-order}
\dot{x}_{\alpha}&=&	i  \frac{g^2 S} {2 x_{\alpha}} + \frac{i \Delta(t)} {2} x_\alpha-i x^3_\alpha\\
&	-&\frac{i g^2}{4}\sum_{\bfrac{\beta=1}{\beta\neq \alpha}}^M \left[ \frac{1}{x_\alpha+x_\beta}+\frac{1}{x_\alpha-x_\beta}\right]\,. \nonumber
\end{eqnarray}

Therefore, it becomes apparent  that the ${x_\alpha}$  move according to a classical Hamiltonian
$H_{I}=\sum_{\alpha=1}^{M}\frac{p^{2}_{\alpha}}{2}+V_\alpha(\{x_{\alpha}\})$, with potential
\beq\label{eq:IMmodel}
V_\alpha(\{x_{\alpha}\})&=&\frac{g^4 }{16} \sum_{\bfrac{\beta=1}{\beta\neq \alpha}}^{M}\left(\frac{1}{(x_{\alpha}-x_{\beta})^{2}}+\frac{1}{(x_{\alpha}+x_{\beta})^{2}}\right)\nonumber\\
&+&\frac{1}{2}x^{6}_{\alpha}-\frac{\Delta(t)}{2} x^{4}_{\alpha}+\frac{\gamma(t)}{2}
x_{\alpha}^{2}+  \frac{g^{4} S^{2}}{8} \frac {1}{x_{\alpha}^{2}} 
\eeq
where
\begin{equation}
\gamma(t)
= ( M-1-S)g^{2} -2 g^2\, S+\frac{\Delta^2}{4}-i \frac{\dot{\Delta}(t)}{2}.
\end{equation}
This model is a complexified version of the BC-type Inozemtsev model
\cite{IM1} and belongs to the family of generalizations of the Calogero model. It is integrable on the classical level for
time-independent parameters. We can therefore interpret the full quantum dynamics and breaking of integrability in terms of the classical equations of motion. 

Here we break the integrability by the time-dependent driving of the detuning. 
Namely, we consider the following setup: at $t=0$ the system
is prepared in its ground state at $\Delta=\Delta_0$. Then we evaluate
numerically its time evolution under the periodic detuning
$ \Delta(t)=\Delta_0\cos(\omega t)$.
We solve the time-dependent Schr\"{o}dinger equation by using a Runge-Kutta integration scheme. 
The rapidities can be obtained from the coefficients of the wave functions by 
finding the roots of symmetric polynomials. For illustration of the principle, 
we choose a small number of excitations, $M=4$, $S=6$ and a strong amplitude of the detuning
$\Delta_0/g=5$, such that the bosonic modes are highly occupied initially,
$N_b=\av{b^\dag b}\approx 3.2$, and the population of excited spins is small.
The high driving amplitude causes strong dynamical redistribution of
excitations between bosonic and spin degrees of freedom.  If the driving
frequency is non-resonant, dynamics remain almost
adiabatic and observables are expected to exhibit periodic oscillations along
the instantaneous ground state values.
\begin{figure}[ht!]
  \begin{center}
    \includegraphics[width=0.51\textwidth]{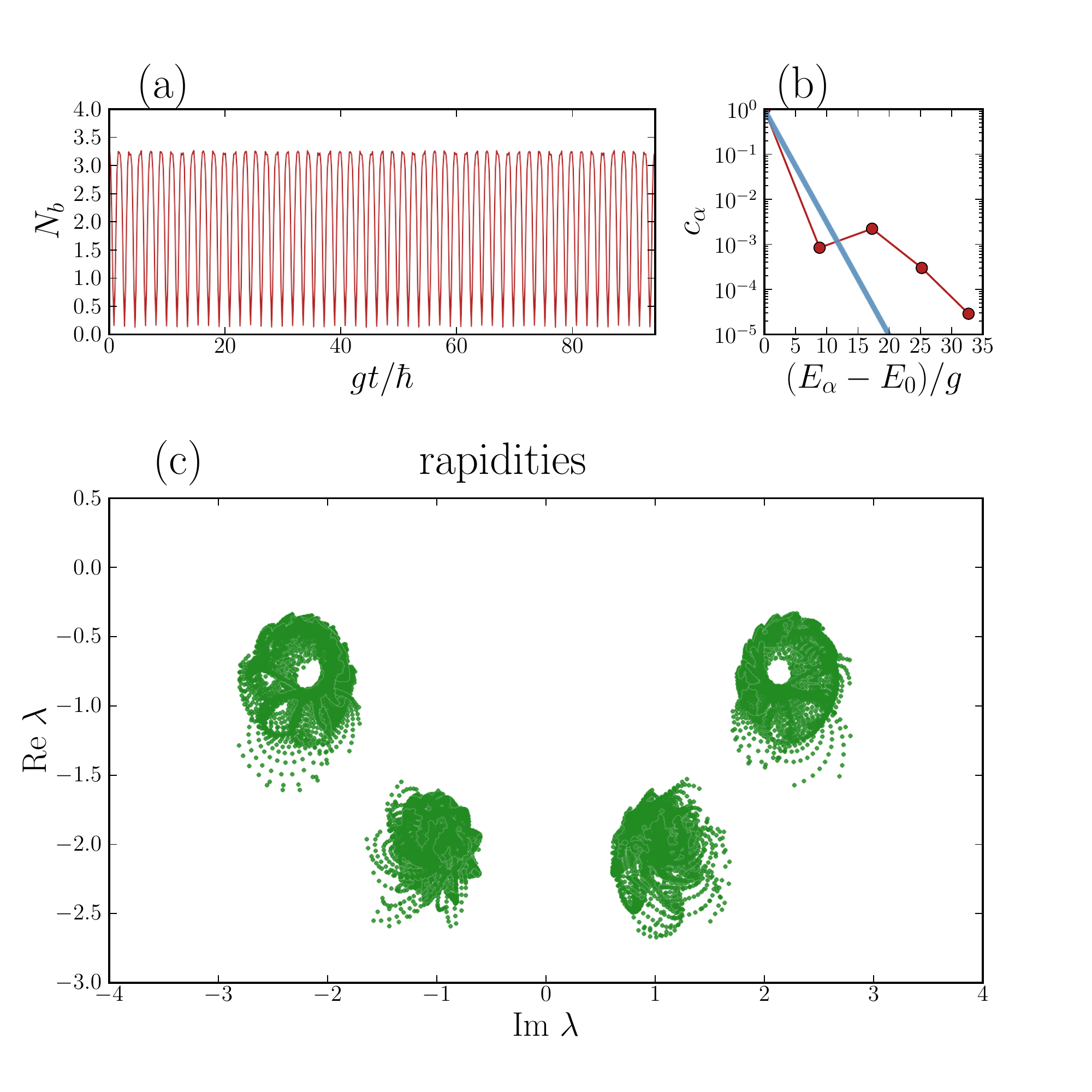}
  \end{center}
  \caption{\label{fig1}
Dynamics of the Tavis-Cummings model driven non-resonantly with the amplitude $\Delta_0/g=5$ and a
frequency $\omega=3.57g/\hbar$, $S=6$ and $M=4$. (a) The boson occupation number
$N_b$ monitored over some interval of time, (b) the weights of eigenstates \eqref{eq:weights}
$c_\alpha$ and (c) the stroboscopic maps of all rapidities $\lambda_m$, $m=1,\dots,M$ after
4000 cycles.}
 \end{figure}

\begin{figure}[ht!]
  \begin{center}
    \includegraphics[width=0.51\textwidth]{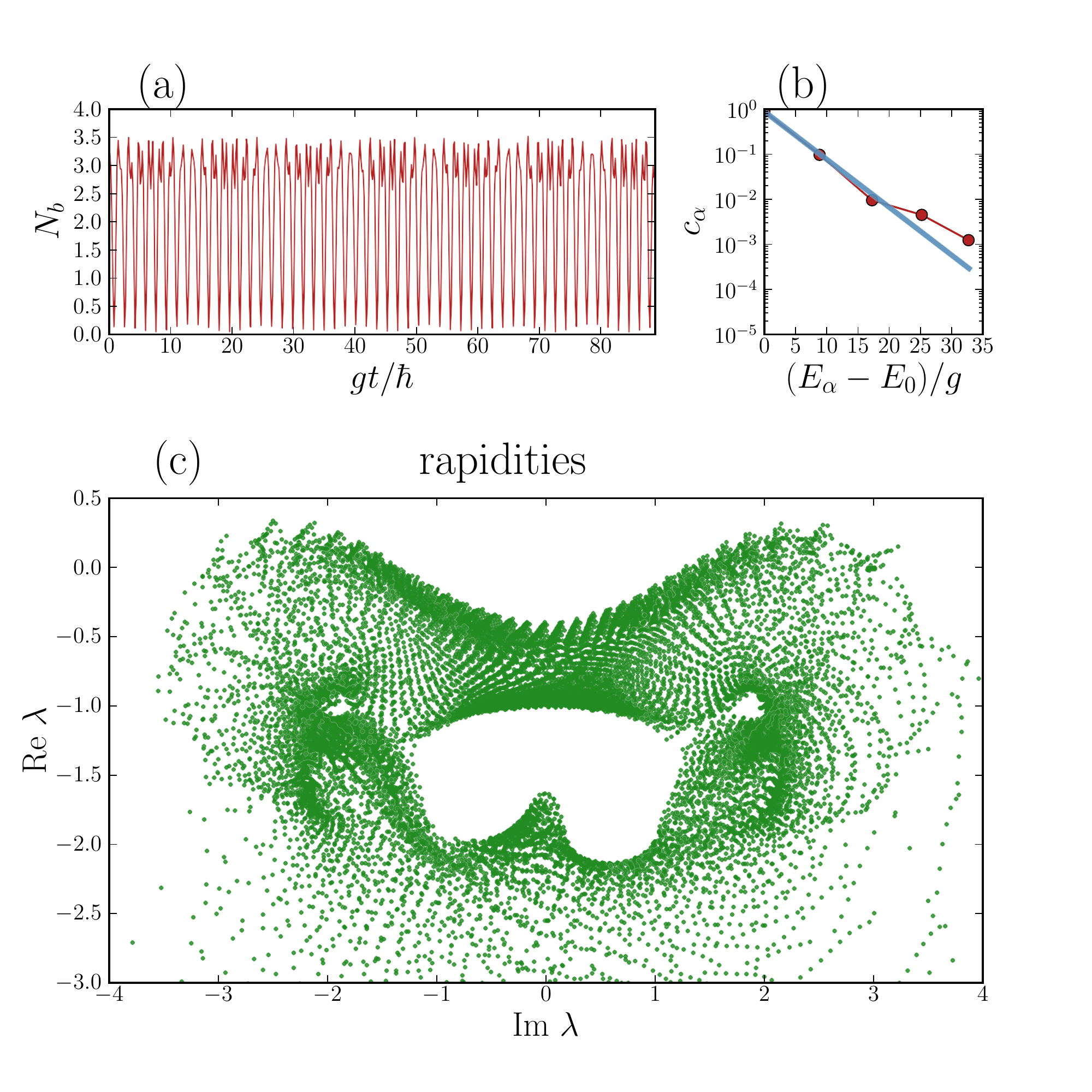}
  \end{center}
  \caption{\label{fig2}
  The Tavis-Cummings model driven near-resonantly with amplitude $\Delta_0/g=5$ and
frequency $\omega=3.68g/\hbar$. For explanations see caption of Fig. 1.}
 \end{figure}
\begin{figure}[t!]
  \begin{center}
   \includegraphics[width=0.51\textwidth]{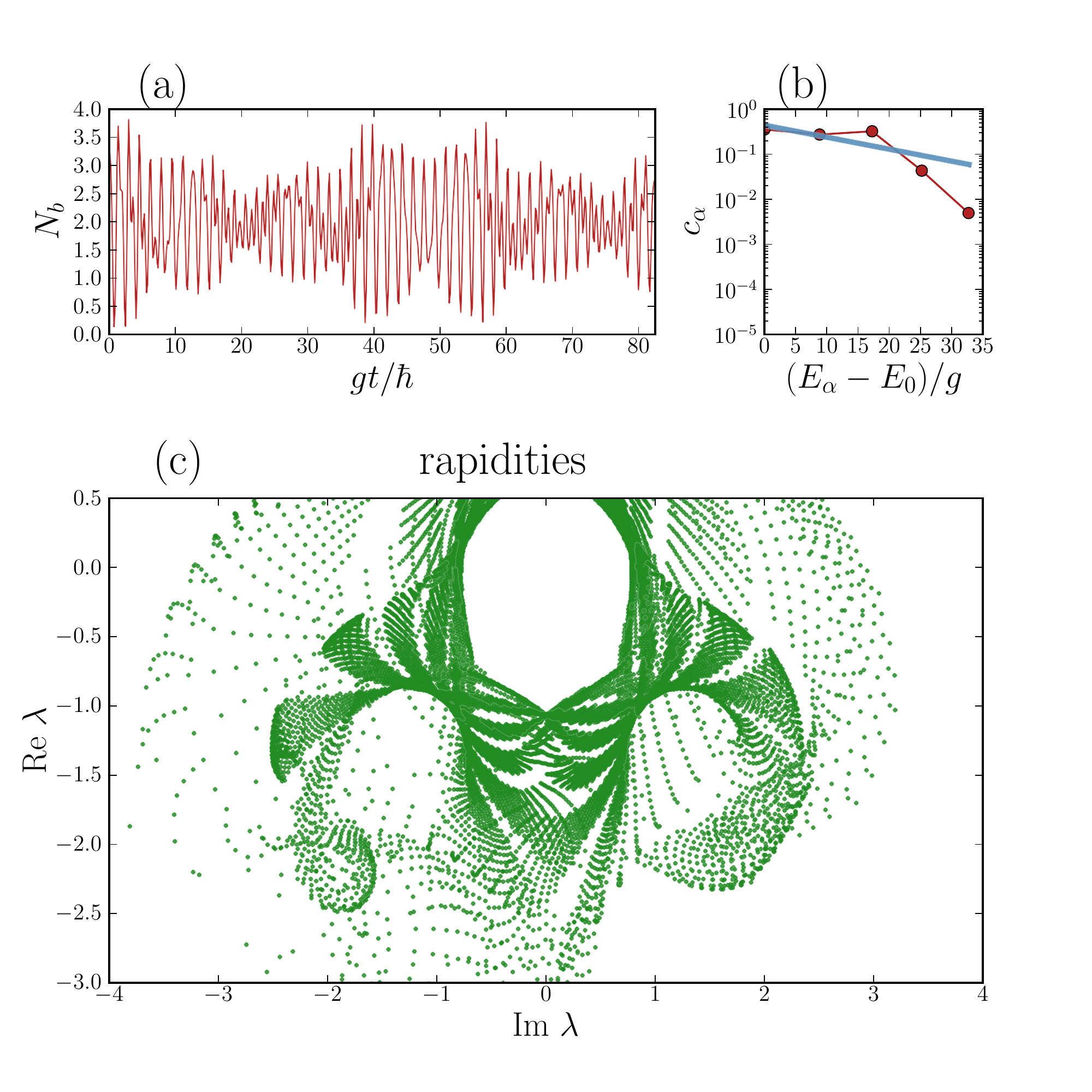}
  \end{center}
  \caption{\label{fig3}
  The Tavis-Cummings model driven with amplitude $\Delta_0/g=5$ and
frequency $\omega=3.75g/\hbar$. For explanations see caption of Fig. 1.}
\end{figure}

In Fig. 1(a) the following example example is shown: At frequency $\omega=3.57 
g/\hbar$, there are regular oscillations of the boson populations $N_b(t)$
between $3.2$ and $0.2$. The rapidities, which correspond to the
position variables of the classical model \eqref{eq:IMmodel}, are monitored
stroboscopically after each cycle (i.e. at time $t_p=2\pi p/\omega$,
$p=0,\dots,P$, where $P=4000$ in the present case) by collapsing them onto a
single complex plane. Fig. 1(c) shows that in this non-resonant case the rapidities cluster on circles located
around the ground state positions. These circles indicate the existence of
stable KAM-tori in the 16-dimensional phase space of the classical system and
according to our correspondence between dynamics of the
quantum and the auxiliary classical system, we can classify such behavior 
as nearly integrable. In order to
characterize the statistical properties of this system we measure the
distribution of states averaged over all cycles
\begin{align}
\label{eq:weights}
  c_\alpha=\frac{1}{P}\sum_p|\braket{\psi(t_p)}{\alpha}|^2\,,
\end{align}
where $\ket{\psi(t_p)}$ is a state of the driven system at $t=t_p$ and
$\ket{\alpha}$ are the eigenstates of the Hamiltonian $H(t_p)$ after $p$
cycles. In Fig. 1(b), we
found their distribution to decay rapidly, which is expected
in this nearly adiabatically driven system.
We compare this distribution to a Boltzmann distribution,
$c_\alpha=e^{-\beta E_\alpha}/Z$, with the same average energy. It turns out that the
Boltzmann distribution cannot describe the weights. Fig. 1(b) also shows that very few energy is pumped into the system.

In Fig. 2 we consider a slight increase of the frequency with respect to
non-resonant case to $\omega=3.68 g/\hbar$. The boson occupation, which starts to
exhibit an additional beating frequency [Fig 2(a)], suggests that a resonance is
approached in the quantum model. Interestingly, this comes along with a
scattering of the rapidities on the collapsed 2-dimensional stroboscopic maps [Fig. 2(c)]. 
From the point of view of the auxiliary classical system, these dynamics rather
strongly deviate from the integrable limit. It has to be noted that despite the relatively dense
exploration of the phase space, we could not find an indication of truly chaotic
behavior. Nevertheless, and despite the small number of degrees of freedom in
the system, this leads to a state distribution remarkably close to Boltzmann
distribution [Fig 2(b)]. 

Further increasing the driving frequency to $\omega=3.75 g/\hbar$ as in
Fig. 3 leads to strongly beating dynamics of boson occupancies. This resonance
of the quantum model leads to a new structured pattern
in the stroboscopic map of the classical variables. This hints that there are new emerging
quasi-periodic orbits, which reside on a topological structure
different from the one of the near-adiabatic case.
The state distribution in Fig. 3(c) shows that the weights deviate considerably from the
Boltzmann distribution. Unlike in the non-resonant case [Fig. 1(c)], a large amount of energy is absorbed by the system.

The cycle structure, well-localized rapidities in the non-resonant cases, and a special pattern in a resonantly driven cases, repeats when further increasing driving frequencies or by modifying other parameters. A special case is the two-particle problem, $M=1$, where a ring-pattern is transformed into a line and back to a ring upon changing the driving frequency.

In summary, we derived a correspondence between a time-dependent quantum model and an auxiliary classical system. The strength of this approach is illustrated by an example of a driven Tavis-Cummings model with a
frequency tuned from a non-resonant to a resonant value.
The emerging dynamics can be interpreted in terms of the classical underlying system,
whose trajectories show very different pattern in their stroboscopic maps. 
At the point where one pattern is deformed into the other, irregularity
in the classical dynamics is most pronounced and time-averages of quantum observables approached thermal equilibrium.

The Tavis-Cummings model belongs to the special class of homogeneous Gaudin models. 
The fact that the rapidities can be used to describe the full quantum dynamics is due to the completeness of the off-shell BA
for the homogeneous model.
Therefore, extending the approach to the inhomogeneous Tavis-Cummings model \cite{Strater2012} or Richardson models \cite{Faribault2009}
is not straightforward. For these models, it is impossible to describe an arbitrary state in terms of a single off-shell Bethe
state with a time-dependent rapidities and a linear superposition of these states would be necessary to capture the dynamics 
of these systems. We believe that the approach based on the separation of variables \cite{Sklyanin} could give an important 
insight into a further development of our approach. 

{\it Acknowledgements --} 
We acknowledge valuable discussions with B. Altshuler, D. Baeriswyl, J.S. Caux,
E. Demler, A. Polkovnikov and A. Tsvelik.  
This work is supported by the Swiss National Science Foundation. PB acknowledges hospitality of the University of Fribourg.

\end{document}